\documentclass[11pt,twoside]{article}

\usepackage{asp2004}
\usepackage{epsf}
\usepackage{lscape}
\usepackage{graphicx}

\begin{document}
\title{Modelling V838 Monocerotis as a {\it Mergeburst} Object}   
\author{Noam Soker}   
\affil{Department of Physics, Technion, Haifa 32000, Israel;
soker@physics.technion.ac.il}
\author{Romuald Tylenda}
\affil{Department for Astrophysics, N.Copernicus Astronomical Center,
Rabia\'nska 8, 87-100 Toru\'n, Poland; tylenda@ncac.torun.pl}

\begin{abstract} 
We discuss the main observational facts on the eruption of V838
Monocerotis  in terms of possible outburst mechanisms. We conclude
that the stellar merger scenario is the only one, which can
consistently explain the observations.
\end{abstract}

\section{Introduction}

The outburst of V838 Monocerotis  was discovered at the beginning
of January 2002 (Brown 2002).
After about a month it suffered another outburst and reached a luminosity
of $\sim 10^6 L_\odot$.
Initially thought to be a nova, the object soon appeared unusual
and enigmatic in its behavior. The eruption, as observed in the optical,
lasted about three months. After developing an A-F supergiant
spectrum at the optical maximum at the beginning
of February 2002, the object showed a general tendency
to evolve to lower effective temperatures. In April 2002
it almost disappeared from the optical but remained very
bright in infrared, becoming one of the coolest M-type
supergiants yet observed. A detailed discussion in Tylenda \& Soker (2006)
shows that novae (or novae-like)
outburst models and born-again asymptotic giant branch (AGB) stars cannot
account for basic properties of V838~Mon.
The merger of two stars, on the main sequence or evolving toward the main
sequence,
emerged as the most promising model.
Our view is that the observations and their analysis (although not
completed)
that were presented at the first meeting devoted to this object
(held in La Palma May 2006), are all in support of the stellar binary merger
model.

Stellar mergers have been recognized for a long time as events which can be
important for evolution of binary systems. Also in discussions of globular
clusters
stellar mergers are usually considered as the most probable source of blue
stragglers
e.g. De Marco et al. (2005, and earlier
references therein; Sills, Adams, \& Davies 2005).
However, the main interest in these cases has been directed towards
understanding
the nature of the final product of a merger in terms of its mass, chemical
structure
and farther evolution. Little attention has usually been paid to direct
observational appearances of these events.
As we think that the appearance itself of merger events will attract more
attention
in the future, we would like to term these type of violent merger events
that lead to V838~Mon type events, {\it mergebursts}.

In this paper written to the proceedings of the meeting, we summarize the
main
properties of V838~Mon that are relevant to its modelling, mainly as they
where
presented or cited during the meeting.
We then summarize some of our results published elsewhere regarding the
merger model.
As we summarize the introduction to the final discussion of the meeting as
presented
by one of us (Soker), we mainly cite talks from the meeting as was done
during the original
presentation, assuming each paper will present the most updated results, and
have a complete
list of relevant references to its subject.
We also hope that all of these papers will be posted on astro-ph, therefore
be
freely available.

\section{Mass}

Despite large amount of observational data, relatively little efforts have
been done to deduce physical properties, in particular the envelope mass,
mass loss,
and mass of echoing material.
We will try to estimate these from the few works that derived
these quantities.

\subsection{The Mass of the Surrounding Material}

Despite several talks on the echo of V838 Mon (Liimets, Kolka, \& Kipper 2006;
Crause 2006; Santander-Garcia \& Corradi 2006; Henden 2006; Bond 2006; Sparks 2006)
no mass estimate was given, beside by Ashok's group.
Ashok (Ashok et al. 2006) presented in his talk an estimate based on IR echo
formed by dust emission. The estimated gas+dust mass from the IR echo is
$M_{\rm echo}({\rm IR})\sim 150 M_\odot$ (Banerjee et al. 2006).
This estimate met some objection from a number of people e.g.,
because the IR echoing is concentrated in a small region at
any given time (Sugerman 2006).
However, it seems quite reasonable, and may fit other estimates.

V838 Mon stayed near its maximum luminosity for a time $\Delta t_o \simeq 2$~months.
Let the echoing gas reside inside a radius $R_{ec} \sim 2$~pc.
Simple estimates show that the echo comes from a volume fraction of
$f_{ec} \sim (c \Delta t_0/R_{ec})\sim 0.03$.
However, the echoing material is not uniform, and it is possible that there was more dust
in the echoing region than the average in the region, say by a factor of $f_u \sim 3$.
It is possible  that there are other IR sources in the field, which cause
diffuse IR emission with low surface brightness,
so the echo accounts for, say, $f_s \sim 0.5$ of the IR emission in the field.
Indeed Afsar \& Bond (2006) claimed for a young open cluster surrounding V838 Mon,
which could heat this dust.
Over all, the corrected IR mass in the volume is
$M_{\rm echo}({\rm corrected-IR}) \sim   M_{\rm echo}({\rm IR}) [(f_s/(f_{ec}f_u)]$,
which for the above values is $M_{\rm echo}({\rm corrected-IR}) \sim 900 M_\odot$.

Sparks (2006) presented some properties of the echo in the visible light. In a private
conversation he estimated the gas particle number density in the echoing matter
to be $n \sim 200~ {\rm cm}^{-3}$.
For an echoing region having radius $R_{ec}$ (even though the echoing region at any time
is not a sphere), and filling factor $f_e$ the total mass from the echo in the visible is
$M_{\rm echo}({\rm visible}) \simeq 100 (R_{ec}/2 {\rm pc})^3 (f_e/0.5) M_\odot$.
This value is lower that the above estimate based on the IR echo measurements of
the Ashok's group.

The results from CO observations are more difficult to reconcile with the above estimates,
and require photo-dissociation of CO molecules.
Kaminski et al. (2006) gave a limit on the total molecular mass of
$\sim 300 M_\odot$ inside a radius of $r=5 {\rm pc}$ around V838 Mon.
This translates to a limit of $\sim 20-50 M_\odot$ inside a radius of $r=2 {\rm pc}$.
Deguchi (2006) has CO observations of five regions, each of
radius 7.5 arcsec, around V838 Mon. He reported the detection of emission in one region,
with an estimated mass of $M_{\rm CO-spot} \simeq 1 M_\odot$.
This translates to a total mass of $\sim (60/7.5)^2M_{\rm CO-spot}/4 \simeq  16 M_\odot$
inside $r \simeq 60~{\rm arcsec} \simeq 2 {\rm pc}$ (assuming a distance
of 6~kpc to V838 Mon; Sparks 2006; Afsar \& Bond 2006).
These mass estimates can be reconciled with the mass estimate from
the visible and IR observations
if most, $\ga 90 \%$, of the CO molecules have been photo-dissociated by
the emission of the surrounding B-stars, which were reported to exist there
(Afsar \& Bond 2006).
Many systems are known where stellar radiation destroys CO molecules while the
warm dust survives, e.g., around A-stars (Greaves et al. 2000).

The conclusion from the $\sim 10^2-10^3 M_\odot$ estimated mass in the echoing region is
that the echoing material is of ISM origin, as suggested in Tylenda (2004)
and  Tylenda, Soker \& Szczerba (2005),
rather than it resulted from previous eruptions as claimed by, e.g., Bond and collaborators
(Bond et al. 2003; Bond 2006).
Even if clear symmetries are seen in the echo images, they simply reflect
the echo geometry.
Also, the progenitor stellar wind could have shaped part of the echoing gas
prior to eruption.
More than that, it is quite likely that the echoing material reveals an axial-symmetric
structure, resulting from the relative motion of the progenitor of V838 Mon through the ISM,
e.g., a bow shock, as discussed in Tylenda (2004) and Tylenda et~al. (2005).

\subsection{The Mass in the Inflated Envelope and Wind}

Many talks during the meeting were devoted to the observation of V838 Mon itself and to
its luminosity and temperature evolution (e.g., reviews by Henden 2006; Munari 2006).
However, the lack of any new attempt to derive physical parameters relevant to
models of V838 Mon was quite disappointing.
In particular, no new mass estimate was given to the inflated envelope and ejected
mass, so we are left with the values given by Tylenda (2005).
The mass lost from the system that was found by Tylenda (2005), translated to the new distance
estimate of 6~kpc, is $M_{\rm lost} \simeq 0.0025 M_\odot$ if a Planck mean opacity is used,
and $M_{\rm lost} \simeq  0.3 M_\odot$ is a Rosseland mean opacity is used.
As well known, the opacity should be closer to the Rosseland mean opacity
(at the meeting this was discussed by Pavlenko 2006).
V838 Mon has strong water absorption lines (Lynch et al. 2006), which indicate that a mass of
$M_{\rm lost} \sim 0.04 M_\odot$ was lost by V838~Mon during the eruption
(Lynch et al. 2004; L. Bernstein private communication 2006).
The conclusion from the above is that the mass lost by V838 Mon is
$M_{\rm lost} \simeq 0.04-0.2 M_\odot$.
We hope more attempts will be made to derive this mass, with the caution that
the envelope might be aspherical (Lane et al. 2005).

The mass of the inflated envelope, i.e. gravitationally contracting in the
decline, is $M_{\rm env} \simeq 0.11 M_\odot$ for a distance of $D= 6$~kpc
(it was scaled from a mass of $0.22 M_\odot$ derived by Tylenda [2005] for the then
assumed distance of $D=8$~kpc).
Over all, the inflated envelope and the expelled mass are added to a total
of extended (i.e. strongly affected) mass of $M_{\rm extended} \simeq 0.1-0.3 M_\odot$.

This extended mass completely rules out any model where a collision of planets with
a star caused the V838 Mon eruption. In principle, a collision of a main sequence
star with a planet can lead to a transient event, but not as bright as that of V838 Mon.
Retter (Retter et al. 2006b) presented the planet capture model of V838 Mon
(Retter \& Marom 2003; Retter et al. 2006a).
The paper by Retter et al. (2006a) suffers from severe problems:
(1) They define a `stopping radius'. This is an ill-defined term.
A planet inside a stellar envelope will not stop at any radius. It will continue to
spiral-in until being evaporated or torn apart by tidal forces, or will overflow
its Roche lobe.
They write: `` . . the impacting planet has come to rest relative to the stellar
envelope, i.e. when it has transferred all (or most of) its
kinetic energy to the parent star. ''
But the kinetic energy of the planets only increases as it spirals-in in a Keplerian orbit.
If the planets really stops, then it will sink inward.
(2) To account for the large energy emitted by V838 Mon they need one or more of the planets
to penetrate deep into the stellar envelope. However, a large amount of energy liberated
deep in an envelope goes mainly to lift the outer envelope gas. When the planet is deep
inside the star the outer envelope mass is much larger than the planet mass, and the envelope
will expand by a very small factor, and with not much increase in luminosity.
(3) They consider either a model with three planets or one. The model with three planets
suffers from that no three stable orbits can be given for the time prior to eruption.
The model with one planet requires the planets to get deep into the envelope, where
most energy will go to slightly lift the outer stellar envelope layers (point 2 above).
(4) From obvious energetic reasons it is impossible to eject more mass than was accreted.
Numerical simulations give that mass lost in a merger is of a few
percent of the accreted star. V838~Mon lost a mass of $M_{\rm lost} \ga 0.04 M_\odot$,
and no planet, nor even a brown dwarf, can account for this.

\section{Other Types of Eruptive Objects That Do Not Fit V838 Mon}

In a previous paper we (Tylenda \& Soker 2006, see the table in that paper) compared the
stellar merger model with models based on thermonuclear run-away eruptions.
It was found that neither models based on born-again AGB stars nor models based on
novae can account for the properties of V838 Mon. In particular,
born-again AGB stars and novae (or nova-like systems) fade as very hot (blue) objects.
This is contrary to the behavior of V838 Mon in the last 4.5 years.
This was the consensus in the meeting. Some people even asked specifically that we stop
calling V838 Mon a nova-like event.

Hirschi (2006) showed that evolution of single massive stars cannot lead to any event
similar to that of V838 Mon.

Lawlor (2006) admitted that his previously suggested model,
based on born again AGB star which accretes mass from a companion, cannot account for
the late behavior of V838 Mon (see also Tylenda \& Soker 2006).

Adding to these the failure of the planet model to account for the energy
and envelope mass of V838 Mon, we are left with the stellar merger model as the only
model which can account for all the observations presented at the meeting.

\section{The Stellar Merger Model}

\subsection{Merger Events: General Considerations}

The discovery of the eruption of V838~Mon in 2002 and subsequent analysis of
its observed evolution, as well as of other similar
objects (V4332 Sgr, M31~RV), have lead us to suggest that these observed events
were likely to be due to stellar mergers (Soker \& Tylenda 2003; Tylenda \& Soker
2006). Likewise, an analysis done in Bally \& Zinnecker (2005) shows that stellar
mergers in cores of young clusters can be source of luminous and spectacular
observational events.

Different processes can lead to stellar mergers. In dense stellar systems direct
collisions of two stars can quite easily happen leading in most cases to a merger.
In multiple star systems dynamical interactions between the components or
encounters between the system and other stars can destabilize stellar trajectories
so that two components collide and merge, and a third star is ejected.

A binary stellar system can lose angular momentum during its evolution, e.g. due to mass loss,
so the separation of the components decreases, which may finally lead to a merger.
In the latter case, the merger is probably often relatively gentle and does not
lead to spectacular events. This happens when the system reaches and keeps
synchronization until the very merger. The relative velocity between the matter
elements from different components is then very low, there is no violent shock heating
and the orbital energy is released on a very long time scale. However when the binary
component mass ratio is low (but not too low for a bright event) the secondary is unable
to maintain the primary in synchronized rotation, so the so-called Darwin instability
sets in and the merger takes place with a large difference between the orbital
velocity of the secondary star and the rotational velocity of the primary envelope.
In this case the merger is expected to be violent, at least in the initial phase
when the large velocity differences are dissipated in shocks.
Soker \& Tylenda (2007) analyze this case.

The gravitational and kinetic energy of the merging binary system
can result in the following observational events:
\begin{enumerate}
\item Flash of light. This flash is formed by emission from a strongly shocked gas,
in the primary and/or secondary envelope, and will be observed as a flash
lasting as long as the secondary is violently slowed-down in the outer regions of
the primary star. This can be from few times the dynamical time of the system
up to a very long time, depending on the condition of the merging system.
For the flash to be bright, the duration should be short, which implies a large
relative velocity between the secondary and rotating primary envelope.
\item Gravitational and kinetic energy of matter expelled to large distances,
and even leaving the system.
The matter that does not leave the system, falls back on a dynamical time scale
at its maximum distance, and when it becomes optically thick it contracts
on its Kelvin-Helmholtz time scale.
The Kelvin-Helmholtz time scale of the inflated envelope is much shorter than
the Kelvin-Helmholtz time scale of the primary, due to the very high luminosity
and very low mass of the inflated envelope.
The large energy in the inflated envelope and its relatively short contraction time
implies that the energy deposited in the inflated envelope results in a bright
phase of the merging system, lasting much longer than the initial flash
(for detail of this process as applied to V383~Mon,
see Tylenda 2005, and Tylenda \& Soker 2006).
\item Gravitational energy of the expanding inner layers of the primary
and/or secondary (even destroying the secondary).
The mass of the layer above the location of the secondary star will be large when the
secondary penetrate the deep layers of the primary star.
When the inner layers of the primary finally relax to
equilibrium it will be on a very long Kelvin-Helmholtz time scale.
\end{enumerate}
Only the first two energy channels are relevant for the formation of bright transient event
from merger. We term these type of events {\it mergebursts}.
The last process will not form any bright event. Therefore, any model where planets penetrate
deep into the primary star, although liberating huge amount of orbital gravitational energy,
cannot account for the eruption of V838~Mon, or of any other bright event.

\subsection{The Merger Model of V838~Mon}

To account for the eruption of V838~Mon we suggest that the merging stars
had the following properties (see Tylenda \& Soker 2006 for detail):
\begin{enumerate}
\item The system was young. This is supported by the B-type primary star
(Tylenda et al. 2005), the claim for a young stellar population in
its vicinity (Afsar \& Bond 2006), and the fact that the echoing matter
is most likely of interstellar origin (as discussed in Sect. 2.1).
\item The primary was a $\sim 8 M_\odot$ main sequence star, as
deduced from observations in Tylenda et~al. (2005).
\item We assume that the secondary star was of mass $M_2 \simeq 0.3 M_\odot$,
which was formed with the primary star.
This low mass implies that the secondary star was a pre-main sequence star at the
time of the merger.
\item We assume that the orbit prior to merger was of very high eccentricity, or
the secondary was deflected toward the primary by multi-stellar interactions.
\item From simple energetic considerations we have found that in order to inflate
a huge envelope, as observed in V838~Mon, the eccentricity should be high,
$e \ga 0.9$, and the low mass companion should be indeed a pre-main sequence star,
such that less energy is required to disrupt it.
\item This has been confirmed by running a ready to use numerical code
(MMAS, version 1.6, package described
in Lombardi et al. 2002, 2003).
These simulations show that the inflated envelope comes almost entirely
from the secondary star.
\item Based on full 3D numerical simulations (e.g., Freitag \& Benz 2005; Dale \& Davies 2006)
we know that merging binary systems with high eccentricity (or
approaching from infinity, which is equal to taking $e=1$), result
in 2-4 episodes of energy dissipation and mass loss ejection. This
is exactly what is inferred for V838~Mon (see a detailed analysis
in Tylenda 2005): one moderate expansion, followed in a month by
ejection of two massive shells separated in time by a few days.
The episodes of mass loss ejection follow the semi periodic variation in
orbital separation. Such an in-and-out separation of the two merging stars is generic
to merger processes.
Even a neutron star starting with a circular orbit around an evolved star
has semi-oscillatory separation variations as it merges with the evolved star
(Terman, Taam, \& Hernquist 1995).
\item It is quite possible that the binary system became interacting before the final merger
which caused the 2002 eruption. This may explain the possible variability of
the V838~Mon progenitor a few years before the outburst (Kimeswenger 2006).
\end{enumerate}

It should be emphasize again that the only theoretical model that existed
before the eruption of V838~Mon, and which could have in principle predicted the
properties of V838~Mon, is the stellar merger model.
All other models (novae; born again AGB stars; models based on planets)
had to {\it invent} specific properties to account for V838~Mon, and even then
failed to explain basic properties of V838~Mon.
The talks and discussions in the meeting at La Palma strengthened this point.

\subsection{Open questions}

The main drawback of our merger model is that it is based
on estimates and approximate considerations. The simple simulations
presented by Tylenda \& Soker (2006; their Sect. 4.2) only demonstrate
that an inflated envelope can be formed from a merger.
Realistic simulations should provide a three-dimensional structure of
the inflated envelope, its evolution after merger, and the expected
light curve.
This would however be a very complex task involving three-dimensional
radiation-hydrodynamics.
What is more secure as far as physics goes, is that main sequence stars accreting
at a high rate can swell to hundreds of solar radii. In addition to previous
papers showing this swelling (cited by Soker \& Tylenda 2003), in his talk Lawlor (2006)
presented similar results when he compared the merger model with the born-again model.
We hope that people who study main-sequence stellar mergers will be invited to
the next meeting on V838 Mon and related objects.

Another open question concerns the statistics. There was no consensus at the meeting
as to which objects should be grouped with V838 Mon (Kimeswenger 2006).
Therefore, the occurrence rate of such events in the galaxy (and beyond) is unknown.
The progenitor of M31-RV was a low mass star (Siegel \& Bond 2006), adding to the potential
population that can form such event. From the theoretical side, if we consider all types
of {\it mergebursts}, i.e., violent mergers (Soker \& Tylenda 2007),
of both massive and low mass stars, a crude estimate leads to a galactic event rate
of once every 10-50 years.

\subsection{Predictions}

The violent merger model for bright transient events, we suggest to call them
{\it mergebursts}, has some predictions for future observations.

(1) As a result of the gravitational contraction of the inflated envelope
V838~Mon will decline in luminosity with a slowly rising effective
temperature. The time scale of the event will increase with time and the
decline will last many decades (perhaps even centuries). No excursion to
very high effective temperatures ($\ga 10^5$~K) will occur.

(2) The inflated envelope has a relatively large specific angular momentum.
As the envelope shrinks, the rotation velocity increases, and when it shrinks to a radius of
$R \la 100 R_\odot$, it will be possible to detect this rotation velocity
$v_{\rm rot} \simeq (10-40) (R/100 R_\odot)^{-1} ~ {\rm km~s}^{-1}$.

(3) A dense dusty disk, resulting from the merger event, might be revealed as the
star shrinks.

(4) Convective motions in the differentially rotating envelope can amplify
magnetic fields. Mass loss would then be an efficient process of angular momentum loss
from the envelope. This would weaken the effects discussed in the above two points.

(5) The echo structure will deviate more from a spherical morphology, resembling more
an ISM cloud. Signs of a bow-shock+tail structure produced by a star blowing wind
and moving through the ISM might be present.

(6) No close, $a \la 1$~AU, companion will be detected around V838 Mon.
     Companions at larger orbital separations are possible. Such a companion
     might have been the perturbing body that cause the merger event.

(7) The merger event could have been caused by a third lower mass star,
$M_3 \la M_2 \simeq 0.3 M_\odot$, that was ejected from the system. Such a star might be
found escaping from V838~Mon with a speed of $\sim 10-100 ~{\rm km~s}^{-1}$.


\acknowledgements 
This paper has partly been supported from the Polish State Committee for Scientific
Research grant no. 2~P03D~002~25.





\end{document}